\documentclass[prl,aps,amsfonts,amssymb,amsmath,floats,floatfix,showpacs,twocolumn,superscriptaddress,unsortedaddress]{revtex4}
\usepackage{graphics}
\usepackage{epsfig,graphicx}

\begin{document}

\title{Crystal-field level inversion in lightly Mn-doped Sr$_3$Ru$_2$O$_7$}

\author{M.A. Hossain}
\affiliation{Department of Physics {\rm {\&}} Astronomy, University of British Columbia, Vancouver, British
Columbia V6T\,1Z1, Canada}
\author{Z. Hu}
\affiliation{II. Physikalisches Institut, Universit\"{a}t zu K\"{o}ln, Z\"{u}lpicher Stra\ss e 77, 50937 K\"{o}ln, Germany}
\author{M.W. Haverkort}
\affiliation{II. Physikalisches Institut, Universit\"{a}t zu K\"{o}ln, Z\"{u}lpicher Stra\ss e 77, 50937 K\"{o}ln, Germany}
\author{T. Burnus}
\affiliation{II. Physikalisches Institut, Universit\"{a}t zu K\"{o}ln, Z\"{u}lpicher Stra\ss e 77, 50937 K\"{o}ln, Germany}
\author{C.F. Chang}
\affiliation{II. Physikalisches Institut, Universit\"{a}t zu K\"{o}ln, Z\"{u}lpicher Stra\ss e 77, 50937 K\"{o}ln, Germany}
\author{S. Klein}
\affiliation{II. Physikalisches Institut, Universit\"{a}t zu K\"{o}ln, Z\"{u}lpicher Stra\ss e 77, 50937 K\"{o}ln, Germany}
\author{J.D. Denlinger}
\affiliation{Advanced Light Source, Lawrence Berkeley National Laboratory, Berkeley, California 94720, USA}
\author{\\H.-J. Lin}
\affiliation{National Synchrotron Radiation Research Center, 101 Hsin-Ann Road, Hsinchu 30077, Taiwan}
\author{C.T. Chen}
\affiliation{National Synchrotron Radiation Research Center, 101 Hsin-Ann Road, Hsinchu 30077, Taiwan}
\author{R. Mathieu}
\affiliation{Department of Applied Physics, University of Tokyo, Tokyo 113-8656, Japan}
\author{Y. Kaneko}
\affiliation{Department of Applied Physics, University of Tokyo, Tokyo 113-8656, Japan}
\author{Y. Tokura}
\affiliation{Department of Applied Physics, University of Tokyo, Tokyo 113-8656, Japan}
\author{S. Satow}
\affiliation{Department of Advanced Materials Science, University of Tokyo, Kashiwa, Chiba 277-8581, Japan}
\author{Y. Yoshida}
\affiliation{National Institute of Advanced Industrial Science and Technology  (AIST), Tsukuba, 305-8568, Japan}
\author{\\H. Takagi}
\affiliation{Department of Advanced Materials Science, University of Tokyo, Kashiwa, Chiba 277-8581, Japan}
\author{A. Tanaka}
\affiliation{Department of Quantum Matter, ADSM, Hiroshima University, Higashi-Hiroshima 739-8530, Japan}
\author{I.S. Elfimov}
\affiliation{Department of Physics {\rm {\&}} Astronomy, University of British Columbia, Vancouver, British Columbia V6T\,1Z1, Canada}
\author{G.A. Sawatzky}
\affiliation{Department of Physics {\rm {\&}} Astronomy, University of British Columbia, Vancouver, British Columbia V6T\,1Z1, Canada}
\author{L.H. Tjeng}
\affiliation{II. Physikalisches Institut, Universit\"{a}t zu K\"{o}ln, Z\"{u}lpicher Stra\ss e 77, 50937 K\"{o}ln, Germany}
\author{A. Damascelli}
\email{damascelli@physics.ubc.ca} \affiliation{Department of Physics {\rm {\&}} Astronomy,
University of British Columbia, Vancouver, British Columbia V6T\,1Z1, Canada}

\begin{abstract}
Sr$_3$(Ru$_{1-x}$Mn$_x$)$_2$O$_7$, in which 4$d$-Ru is substituted by the more localized 3$d$-Mn, is studied by x-ray dichroism and
spin-resolved density functional theory. We find that Mn impurities do not exhibit the same 4+ valence of Ru, but act as 3+ acceptors; the extra
$e_g$ electron occupies the in-plane 3$d_{x^2\!-\!y^2}$ orbital instead of the expected out-of-plane 3$d_{3z^2\!-\!r^2}$. We propose that the
$3d\!-\!4d$ interplay, via the ligand oxygen orbitals, is responsible for this crystal-field level inversion and the material's transition to an
antiferromagnetic, possibly orbitally-ordered, low-temperature state.
\end{abstract}

\date{Received \today}


\maketitle

Ruthenium oxides are a particularly interesting class of materials exhibiting Fermi liquid properties,
unconventional superconductivity, ferromagnetism and metamagnetism, antiferromagnetic insulating behavior,
and orbital ordering. The richness of this physics is a testament of the intimate interplay between charge,
spin, orbital, and lattice degrees of freedom, despite the fact that 4$d$ transition metals (TM) are not
considered to induce strong electronic correlations. The radial extend of the 4$d$ wave functions is
significantly larger than for TM-3$d$ and even O-2$p$ orbitals. This leads to, on the one end, weaker
correlation effects than in 3$d$ TM-oxides; on the other hand, to an interesting competition between local
and itinerant physics. Doping a TM\,4$d$-O\,2$p$ host with dilute 3$d$ TM impurities might thus be extremely
effective in tuning valence, spin, and, orbital characteristics and, in turn, the macroscopic physical
properties.
\begin{figure}[b]
\centerline{\epsfig{figure=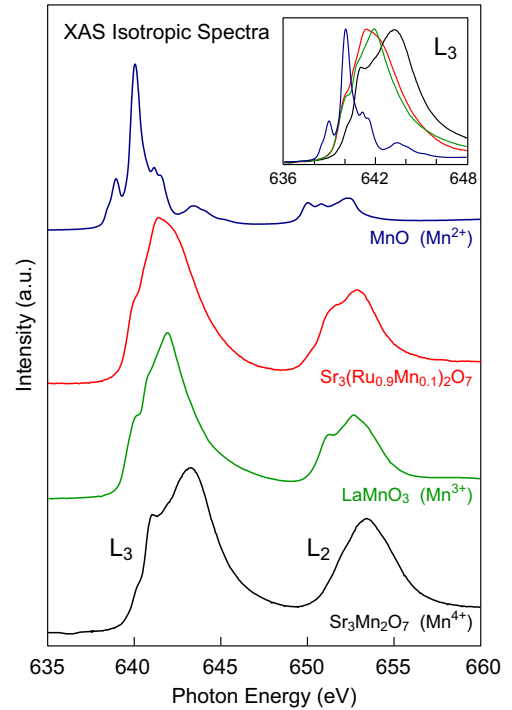,width=0.75\linewidth,clip=}} \caption{(color online). Isotropic Mn $L_{2,3}$-edge XAS data from
Sr$_3$(Ru$_{0.9}$Mn$_{0.1}$)$_2$O$_7$ and stoichiometric Mn-oxides of known valences. Inset: detailed view of the $L_3$-edge chemical
shift.}\label{fig1}
\end{figure}
\begin{figure*}[t!]
\centerline{\epsfig{figure=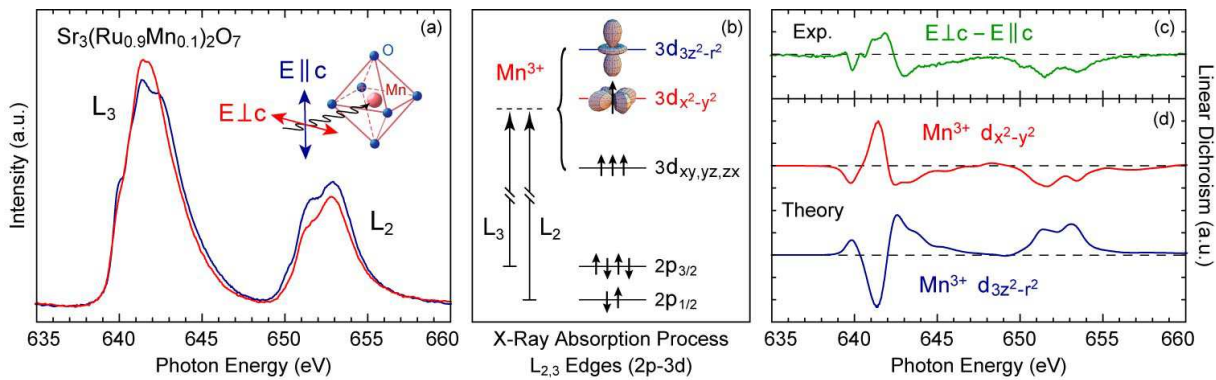,width=0.90\linewidth,clip=}} \caption{(color online). (a) Polarization-dependent Mn
$L_{2,3}$-edge XAS spectra from Sr$_3$(Ru$_{0.9}$Mn$_{0.1}$)$_2$O$_7$ at $T\!=\!295$\,K. (b) Scheme of the XAS process: the $L_2$ ($L_3$) edge
corresponds to the excitation of a Mn 2$p_{1/2}$ (2$p_{3/2}$) electron to the Mn 3$d$ valence shell. The $L_3$-$L_2$ energy separation is due to
the 2$p$ core level spin-orbit coupling. (c) Corresponding experimental linear dichroism: LD$=\![I_{XAS}({\bf E}\!\perp\!c)\!-\!I_{XAS}({\bf
E}\!\parallel\!c)]$. (d) Calculated LD spectra for two possible $e_g$-orbital occupations $(x\!\parallel\! a, y\!\parallel\! b, z\!\parallel\!
c)$.} \label{XAS_comp_295K}
\end{figure*}
We will explore these ideas in the hotly debated Sr$_3$(Ru$_{1-x}$Mn$_x$)$_2$O$_7$ family: metamagnetism and a field tuned quantum phase
transition were discovered on the pure compound \cite{perry_1,perry_2,grigera_1}; with the inclusion of a few percent of Mn, a metal-insulator
transition was observed in transport experiments and the emergence of a Mott-like antiferromagnetic state was proposed \cite{mathieu}.

To address the nature of the metal-insulator phase transition in Sr$_3$(Ru$_{1-x}$Mn$_x$)$_2$O$_7$ and the role of Mn impurities, we used x-ray
absorption spectroscopy (XAS). This is an element and site-specific probe that provides information on the electronic structure of impurities
and host atoms separately, as well as on the emergence of magnetic correlations. By performing polarization-dependent XAS experiments, one can
study linear dichroism (LD) and magnetic circular dichroism (MCD). While MCD in an externally applied magnetic field is sensitive to the
expectation value of the local magnetic moment $\langle {\bf M} \rangle$ and is thus a probe of ferromagnetism, LD is proportional to $\langle
{\bf M}^2 \rangle$ and in turn the nearest-neighbor spin-spin correlation function, thus providing unique information on orbital population
\cite{chen,huang} as well as antiferromagnetic order \cite{kuiper, alders}. We performed temperature-dependent MCD experiments in a 0.5 Tesla
field on Mn-doped Sr$_3$Ru$_2$O$_7$ for Mn concentration as high as 20\%, without detecting any signal above the 2\% noise level, which excludes
ferromagnetism down to 15K. In this paper, we will thus concentrate on room-temperature XAS-LD experiments.

Single crystals of Sr$_3$(Ru$_{1-x}$Mn$_x$)$_2$O$_7$ were grown by the floating zone technique \cite{mathieu}. Total electron-yield XAS
measurements were performed at the Dragon beamline at NSRRC in Taiwan and at beamline 8.0.1 at ALS in Berkeley (the energy resolution was 0.3 eV
and the degree of linear polarization $\sim\!98$\%). All sample surfaces were prepared by in-situ cleaving at pressures better than
$2\!\times\!10^{-9}$\,mbar. The XAS data were normalized to the beam intensity $I_0$; the absolute energy calibration (with accuracy
$\sim\!0.02$\,eV) was obtained from the simultaneous XAS measurement of a MnO single crystal performed in a separate chamber with a small part
of the beam \cite{kurata}.

Before discussing the LD results and possible ordering phenomena, we will address the very basic question of what is the valence of Mn in
Sr$_3$(Ru$_{1-x}$Mn$_x$)$_2$O$_7$. The pure compound is ionic with valence Sr$^{2+}$, Ru$^{4+}$, and O$^{2-}$, which would suggest the
substitution of Ru$^{4+}$ with Mn$^{4+}$ upon doping. Interestingly, in the related compounds SrRu$_{1-x}$Cr$_x$O$_3$ \cite{attfield} and
CaRu$_{1-x}$Cr$_x$O$_3$ \cite{cao} one might also expect the naive Ru$^{4+}\!\rightarrow$Cr$^{4+}$ substitution. However, although the valence
was not probed directly, from the anomalous dependence of the unit cell volume on Cr content a substantial
Ru$^{4+}$+Cr$^{4+}\!\rightarrow$Ru$^{5+}$+Cr$^{3+}$ charge transfer was proposed \cite{attfield,cao}. Thus, in the present case, the 4+ valence
of Mn should not be taken for granted but determined experimentally. Room-temperature, isotropic Mn $L_{2,3}$-edge XAS data from 10\% Mn-doped
Sr$_3$Ru$_2$O$_7$ are presented in Fig.\,\ref{fig1}, together with the results from other Mn-oxide compounds characterized by a well-defined Mn
valence, such as MnO (2+), LaMnO$_3$ (3+), and Sr$_3$Mn$_2$O$_7$ (4+) (see Fig.\,\ref{XAS_comp_295K}b for a description of the $L_{2,3}$-edge
XAS process). As shown in Fig.\,\ref{fig1}, the energy position of the $L_{2,3}$ absorption edge is exquisitely sensitive to the valence of an
element. The shift of the center of gravity to high energy upon increasing the Mn valence from 2+ to 4+ (i.e., `chemical shift'), and the very
close match between Sr$_3$(Ru$_{0.9}$Mn$_{0.1}$)$_2$O$_7$ and LaMnO$_3$ $L_{2,3}$-edge energy and lineshape (inset of
Fig.\,\ref{XAS_comp_295K}), provide already the first surprise: Mn impurities in Sr$_3$Ru$_2$O$_7$ act as Mn$^{3+}$ electron acceptors.

\begin{figure}[t]
\centerline{\epsfig{figure=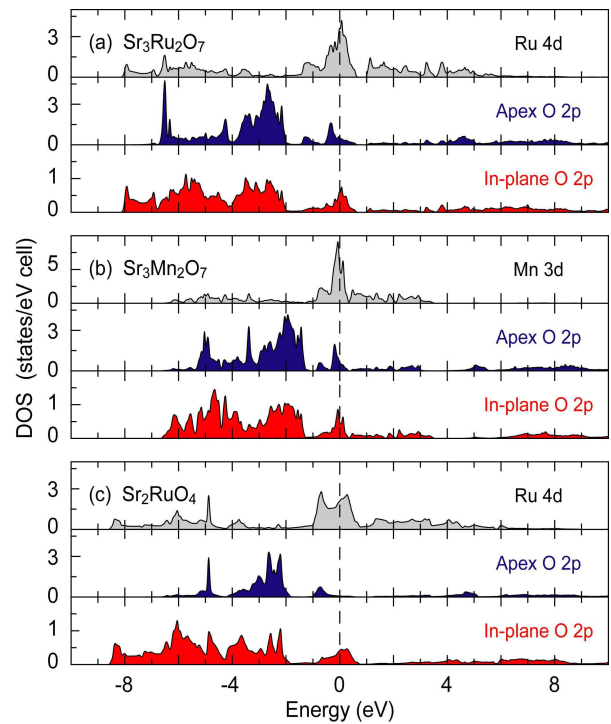,width=0.92\linewidth,clip=}} \caption{(color online). Density-of-states (DOS) of stoichiometric
(a) Sr$_3$Ru$_2$O$_7$ (Ru327), (b) Sr$_3$Mn$_2$O$_7$ (Mn327) with the crystal structure of Ru327, and (c) Sr$_2$RuO$_4$ (Ru214).}\label{LDA}
\end{figure}

\begin{figure}[t]
\centerline{\epsfig{figure=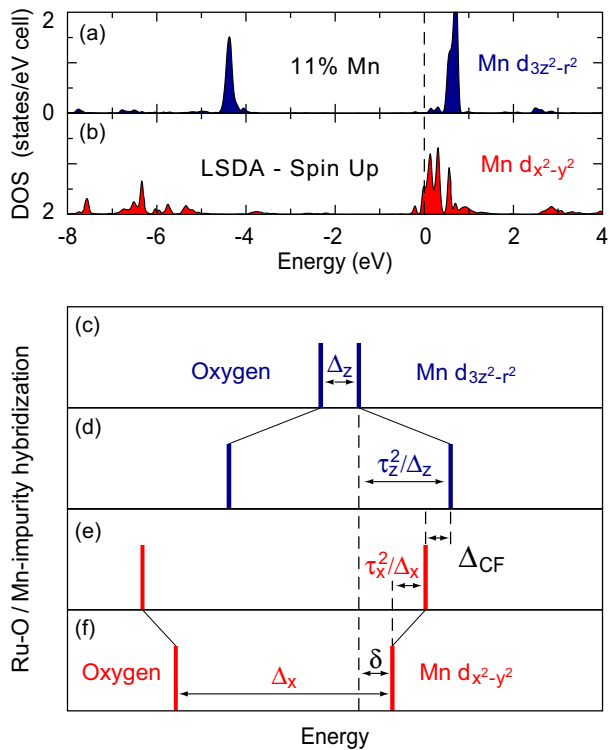,width=0.92\linewidth,clip=}} \caption{(color online). (a,b) Spin-up density-of-states (DOS) from
LSDA calculations for out-of-plane $d_{3z^2\!-\!r^2}$ and in-plane $d_{x^2-y^2}$ Mn $e_g$ orbitals in 11\% Mn-doped Ru oxides. (c-f)
Ru-O/Mn-impurity hybridization leading to the Mn orbital hierarchy inversion: (c,f) energy of out-of-plane and in-plane Mn-impurity and O
orbitals in the Sr$_3$Ru$_2$O$_7$ host material before the Mn-O hybridization is turned on; (d,e) once the O/Mn-impurity hybridization is
considered, the Mn $e_g$-orbital hierarchy is inverted ($\Delta_{CF}\!>\!0$) if $\tau_z^2 / \Delta_z\!-\!\tau_x^2 /
\Delta_x\!>\!\delta$.}\label{valence_U}
\end{figure}

The 3+ valence of Mn has important consequences. While Mn$^{4+}$ has three $d$ electrons in the 1/2-filled $t_{2g}$ shell, Mn$^{3+}$ has an
extra $e_g$ electron ($t_{2g}^3e_g^1$) and is Jahn-Teller active, which adds the orbital dimension to the problem. The key question is whether
the $e_g$ electron will occupy the in-plane $d_{x^2\!-\!y^2}$ or out-of-plane $d_{3z^2\!-\!r^2}$ orbital. Since Sr$_3$Ru$_2$O$_7$ is a
tetragonally distorted system with RuO$_6$ octahedra elongated along the $c$-axis, crystal-field splitting would lead to the occupation of the
out-of-plane $d_{3z^2-r^2}$ orbital. This intuitive picture can be directly verified by x-ray LD. Room-temperature, linearly polarized Mn
$L_{2,3}$-edge XAS spectra from 10\% Mn-doped Sr$_3$Ru$_2$O$_7$ are presented in Fig.\,\ref{XAS_comp_295K}a. The corresponding LD, defined as
the difference between XAS spectra acquired with light polarization perpendicular (${\bf E}\!\perp\!c$) and parallel (${\bf E}\!\parallel\!c$)
to the crystal $c$ axis, is shown in Fig.\,\ref{XAS_comp_295K}c. Information about the orbital population of Mn $e_g$  levels can be extracted
through the detailed comparison between the measured LD and multiplet cluster calculations for different electronic configurations of the Mn,
which are shown in Fig.\,\ref{XAS_comp_295K}d. As we will discuss below, the multiplet calculations are based on parameters from our {\it
ab-initio} density functional theory results (Fig.\,\ref{LDA} and \ref{valence_U}), and are thus not an arbitrary fit of the data. The two
simulated LD spectra are opposite to each other in terms of sign and reveal the in-plane $d_{x^2-y^2}$ orbital polarization for the Mn$^{3+}$
$e_g$ electrons (Fig.\,\ref{XAS_comp_295K}b), instead of the expected $d_{3z^2-r^2}$. Also, the room-temperature LD spectra are virtually
indistinguishable for all of the Sr$_3$(Ru$_{1-x}$Mn$_x$)$_2$O$_7$ samples we studied (not shown), indicating that the Mn$^{3+}$ valence and
$d_{x^2-y^2}$ orbital polarization persist across the whole 5-20\% doping range.

This {\it inversion} of the conventional {\it crystal-field orbital hierarchy} at Mn impurities in Sr$_3$(Ru$_{1-x}$Mn$_x$)$_2$O$_7$ is a very
surprising result. To illustrate this point, we have performed local-density approximation (LDA) band-structure calculations for undoped
Sr$_3$Ru$_2$O$_7$ (Ru327) and Sr$_3$Mn$_2$O$_7$ (Mn327), with the full-potential linearized augmented plane-wave density functional theory code
WIEN2K. In both cases we have used structural data for Ru327 \cite{shaked??-jssc-00}, since our end goal will be that of studying Ru327 for
dilute Ru-Mn substitution. As shown in Fig.\,\ref{LDA}a and b, the basic electronic structures for the two stoichiometric compounds is very
similar and is set by the overlap of O-2$p$ and TM-$d$ orbitals. However, the Mn-3$d$ orbitals are more spatially localized than the rather
extended Ru-4$d$ orbitals; this leads to a reduced bonding-antibonding splitting with the O-2$p$ states and, as clearly evidenced by the
in-plane O-2$p$ density-of-states (DOS), to an overall 30\% bandwidth reduction in Mn327 as compared to Ru327. As for the states of mainly
TM-$d$ character, the $t_{2g}$ band is located within a 1-2\,eV range about the chemical potential and is partially occupied (corresponding to a
4+ valence for both Mn and Ru), while the $e_g$ states are higher in energy and completely unoccupied. Most importantly, as summarized in
Table\,\ref{moments}, the first moments of the TM-$e_g$ partial DOS indicate a lower energy for the $d_{3z^2-r^2}$ than $d_{x^2-y^2}$ states in
both Ru327 and Mn327, consistent with the standard crystal field description for elongated TM-O$_6$ octahedra.

To understand the origin of the level inversion observed in Sr$_3$(Ru$_{1-x}$Mn$_x$)$_2$O$_7$, we have to
further our density functional theory study with the inclusion of dilute Mn impurities. The close similarity
of Ru327 and Sr$_2$RuO$_4$ (Ru214) electronic structures, with almost identical bandwidths
(Fig.\,\ref{LDA}a,c), first moments (Table\,\ref{moments}), and marginal $k_z$ dispersion (not-shown),
suggests that the computationally demanding problem of performing calculations for dilute impurities in
bilayer Ru327 can be more efficiently solved in single-layer Ru214 \cite{chmaissem-prb-98} (the main
difference between the two is the apical oxygen DOS in Fig.\,\ref{LDA}a,c, which is due to the presence of an
additional apical site within the RuO$_2$ bilayer in Ru327 and is not relevant to the present discussion).
\begin{table}[b]
\begin{center}
\begin{tabular*}{8cm}{@{\extracolsep{\fill}}cccc}\hline\hline
Orbital & Ru327 & Mn327 & Ru214  \\
       Symmetry       &(eV) & (eV)  & (eV) \\
\hline
 \!\!TM$_{x^2-y^2}$   &\,\,3.23 &\,\,1.94 &\,\,3.20 \\
 TM$_{3z^2-r^2}$  &\,\,2.93 &\,\,1.82 &\,\,2.93 \\
 MO$_{3z^2-r^2}$  &-6.00 &-4.20 &-5.73 \\
 \!MO$_{x^2-y^2}$   &-6.95 &-4.94 &-7.38 \\
\hline \hline
\end{tabular*}
\end{center}
\caption{DOS first moments ($\int \omega\,DOS(\omega)\,d\omega$) calculated for oxygen molecular orbitals
(MO) and transition metal (TM) $e_g$ orbitals with $x^2-y^2$ and $3z^2-r^2$ symmetry, for stoichiometric
Ru327, Mn327 (in the Ru327 structure), and Ru214. Negative (positive) values identified occupied (unoccupied)
states.} \label{moments}
\end{table}
We have thus performed spin-polarized calculations (LSDA) for Mn-doped Ru214 with a 3x3 supercell in the $ab$-plane ($\sim\!11$\% Mn).

As shown by the spin-up $e_g$ Mn DOS in Fig.\,\ref{valence_U}a,b, in which Mn-$d_{x^2-y^2}$ is lower in
energy than Mn-$d_{3z^2\!-\!r^2}$, LSDA calculations do reproduce the crystal-field inversion discovered in
Sr$_3$(Ru$_{1-x}$Mn$_x$)$_2$O$_7$. This result, which as we have seen is specific to the case in which Mn is
introduced as an impurity, originates from the interplay between the spatially confined Mn-3$d$ orbitals and
the very extended, yet very anisotropic, Ru4$d$-O2$p$ electronic backbone of the Ru-O host. Let us illustrate
this unusual behavior on the basis of the qualitative hybridization scheme of Fig.\,\ref{valence_U}c-f.
Before the O/Mn-impurity hybridization is turned on (Fig.\,\ref{valence_U}c and \,f), the Mn $e_g$ orbitals
are arranged according to the conventional crystal-field splitting for elongated TM-O$_6$ octahedra, with a
small positive difference $\delta\!=\!\epsilon_x\!-\!\epsilon_z$ for the on-site energies $\epsilon_{x,z}$ of
the Mn $d_{x^2-y^2}$ and $d_{3z^2-r^2}$ orbitals. As the hybridization is turned on, however, a level
inversion $\Delta_{CF}\!>\!0$ can be realized if $\tau_z^2 / \Delta_z\!-\!\tau_x^2 / \Delta_x\!>\!\delta$,
where $\Delta_x$ ($\Delta_z$) is the on-site energy difference between Mn $d_{x^2-y^2}$ ($d_{3z^2-r^2}$) and
in-plane (apical) oxygen ligand orbitals, and $\tau_x$ ( $\tau_z$) is the in-plane (out-of-plane)
O-ligand/Mn-$d$ hybridization parameter. Since $\tau_x\!>\!\tau_z$ for elongated octahedra, the level
inversion requires $\Delta_x\!\gg\!\Delta_z$, which is indeed realized in Mn-doped Ru214 and Ru327 with a few
eV difference between $\Delta_{x}$ and $\Delta_{z}$. Note, however, that although the LSDA calculations of
Fig.\,\ref{valence_U}a,b do capture the orbital hierarchy revealed by XAS-LD experiments, the location of the
Mn $d_{x^2-y^2}$ orbitals right above the chemical potential would still lead to a Mn$^{4+}$ valence. This
can be understood as a consequence of the self-interaction that in LDA reduces the ionization energy of
localized states more than that of extended states, making it easier for Mn impurities embedded in the Ru-O
host to achieve a higher oxidation state. To some extent the LDA+U scheme takes care of the self-interaction
problem and Mn$^{3+}$ is obtained in calculations with $U_{Mn}\!>4$\,eV \cite{LDAU}.

The findings of our density functional theory study provide an ab-initio foundation for the ligand field
calculations of the Mn multiplet electronic structure and LD spectra presented in Fig.\,\ref{XAS_comp_295K}d.
These were performed with the program XTLS8.3 \cite{Tanaka94} and the parameters $U_{dd}$=4.5, $U_{dp}$=6.0,
$\Delta$={0.5}, $\delta\!=\!\epsilon_x\!-\!\epsilon_z\!=\!0.2$, $\tau_{x(z)}=\sqrt{3}pd\sigma_{x(z)}$, all
expressed in eV, where: $U_{dd}$ and $U_{dp}$ are the Mn $d\!-\!d$ and Mn $d\!-\!2p$ average Coulomb
repulsion; $\Delta$ is the multiplet average energy difference between the Mn\,$d^4$ and the
Mn\,$d^5\underline{L}$ configuration; $pd\sigma_z\!=\!-1.2$\,eV is the out-of plane O-ligand/Mn-$d$
hopping-integral, which has been taken to be 80\% of the in-plane $pd\sigma_x\!=\!-1.5$\,eV. Finally, the
$d_{3z^2-r^2}$ LD spectrum in Fig.\,\ref{XAS_comp_295K}d was calculated with degenerate ligand orbitals, in
which case $d_{3z^2-r^2}$ is occupied. The $d_{x^2-y^2}$ LD spectrum is instead obtained with
$\Delta_x\!-\!\Delta_z$=4.7\,eV, and the close agrement with the experimental LD spectrum
(Fig.\,\ref{XAS_comp_295K}c,d) provides a direct confirmation of our analysis.

We have shown that, at the microscopic level, the remarkable sensitivity of Sr$_3$Ru$_2$O$_7$ to light Mn
doping stems from the interplay of the localized Mn\,3$d$ impurity with extended Ru\,4$d$-O\,2$p$ orbitals.
Clearly, one should include in the calculations also detailed information on the local structure around the
Mn impurities: a contraction along the $c$ axis and elongation along $a$ and $b$ axes have been seen by x-ray
diffraction across the metal-insulator transition \cite{mathieu}, which might imply a compression of the
MnO$_6$ octahedra. Our study indicates that, although such compression would be consistent with - and further
enhance - the observed crystal-field level inversion, the driving mechanism for this phenomenon is purely of
electronic origin. More generally, the substitution of 3$d$-TM impurities in 4$d$ and even 5$d$ TM-oxides
might provide a novel, powerful approach to the tailoring of the physical properties of complex electronic
oxides.

This work is supported by the Sloan Foundation (AD), ALS Doctoral Fellowship (MAH) and CRC (AD, GAS)
Programs, NSERC, CFI, CIFAR, and BCSI. ALS is supported by the U.S. DOE under Contract No. DE-AC02-05CH11231,
and the research in K\"oln by the Deutsche Forschungsgemeinschaft (DFG) through SFB 608.

\bibliographystyle{plain}

\end{document}